\documentclass[twoside]{article}
\usepackage{fleqn,espcrc2}
\usepackage{graphicx}
\def\bef{\begin{figure}}
\def\eef{\end{figure}}

\newcommand{\be}[1]{\begin{equation}\label{#1}}
\newcommand{\beq}{\begin{equation}}
\newcommand{\ee}{\end{equation}}
\newcommand{\beqn}[1]{\begin{eqnarray}\label{#1}}
\newcommand{\eeqn}{\end{eqnarray}}
\newcommand{\bd}{\begin{displaymath}}
\newcommand{\ed}{\end{displaymath}}
\newcommand{\mat}[4]{\left(\begin{array}{cc}{#1}&{#2}\\{#3}&{#4}
\end{array}\right)}
\newcommand{\matr}[9]{\left(\begin{array}{ccc}{#1}&{#2}&{#3}\\
{#4}&{#5}&{#6}\\{#7}&{#8}&{#9}\end{array}\right)}

\def\lsim{\raise0.3ex\hbox{$\;<$\kern-0.75em\raise-1.1ex
\hbox{$\sim\;$}}}
\def\gsim{\raise0.3ex\hbox{$\;>$\kern-0.75em\raise-1.1ex
\hbox{$\sim\;$}}}
\def\simlt{\mathrel{\lower2.5pt\vbox{\lineskip=0pt\baselineskip=0pt
           \hbox{$<$}\hbox{$\sim$}}}}
\def\simgt{\mathrel{\lower2.5pt\vbox{\lineskip=0pt\baselineskip=0pt
           \hbox{$>$}\hbox{$\sim$}}}}
\def\unity{{\hbox{1\kern-.8mm l}}}
\def\16p{16\pi^2}
\newcommand{\eps}{\varepsilon}

\def\la{\lambda}

\newcommand{\ov}{\overline}
\renewcommand{\to}{\rightarrow}

\def\bY{{\mathbf Y}}

\def\bX{{\mathbf X}}
\def\bS{{\mathbf S}}
\def\bM{{\mathbf M}}
\def\bA{{\mathbf A}}

\def\bG{{\mathbf G}}
\def\bP{{\mathbf P}}

\def\bb{{\mathbf b}}

\def\bmu{\mathbf m }

\def\cA{{\cal A}}

\def\cL{{\cal L}}

\def\cR{{\cal R}}
\def\cS{{\cal S}}

\newcommand{\AmS}{{\protect\the\textfont2
  A\kern-.1667em\lower.5ex\hbox{M}\kern-.125emS}}

\title{\vskip -0.2cm 
\hfill{\small DFAQ-01/TH/0}\\
\vskip -0.2cm 
\hfill{\small DFPD-01/TH/23}\\
\vspace{ 0.4cm}
Flavour structure,  flavour symmetry and supersymmetry 
\thanks{Based on the talk given by Z. Berezhiani at the Int. Workshop 
`30 Years of Supersymmetry', Minneapolis, USA, 13-27 Oct. 2000}
}

\author{Zurab Berezhiani\address{
Universit\`a di L'Aquila, 
I-67010 Coppito AQ, and \\
INFN, Laboratori Nazionali del Gran Sasso, I-67010 Assergi AQ, Italy} 
and
Anna Rossi\address{ Universit\`a di Padova and 
INFN Sezione di Padova, I-35131 Padova, Italy
}}

\begin{document}

\begin{abstract}
We discuss the role played by the horizontal flavour symmetry 
in supersymmetric theories. 
In particular, we consider the horizontal symmetry $SU(3)_H$ 
between the three fermion families and show how this 
concept can help in explaining the fermion mass spectrum 
and  mixing pattern in the context of SUSY GUTs.

%\thanks{
%}. 

\end{abstract}

\maketitle

\section{Introduction}
One of the most obscure sides of  particle physics concerns 
the fermion flavor structure. This is a complex problem  
with different aspects 
questioning the origin of the mass spectrum and mixing pattern 
of quarks and leptons (including neutrinos) and CP-violation, as well as 
the suppression of  flavor changing (FC) neutral currents, 
the strong CP-problem, etc. \cite{ICTP}. 
Presently, thanks to the new data from the atmospheric and 
solar neutrino experiments, 
the flavor problem is getting more intriguing. 
On the one hand, the experimental data hint to a hierarchical 
neutrino mass spectrum,  similarly to the case of   
the charged leptons and quarks. On the other hand, the lepton 
mixing pattern strongly differs from that of the quarks.  
In particular, 
the 2-3 lepton mixing angle  is nearly maximal, 
$\theta^l_{23} \simeq 45^\circ$ in contrast with 
the analogous quark mixing angle, 
$\theta^q_{23} \simeq 2^\circ$.

The concept of supersymmetry {\it per se} does not help 
in understanding the fermion flavor structure,  
and in addition it creates another 
problem, the so called supersymmetric flavor problem,   
related to the sfermion mass and mixing pattern.  

In the MSSM the fermion sector consists of chiral superfields 
containing the quark and lepton species of three families:  
$q_i=(u,d)_i$, $\bar{u}_i$, $\bar{d}_i$, 
$l_i=(\nu,e)_i$  and $\bar{e}_i$ ($i=1,2,3$). 
The charged fermion masses emerge from the Yukawa terms 
in the superpotential: 
\be{Yuk}
W_{\rm Yuk} = \bY_u^{ij}\bar{u}_i q_j H_2 + 
\bY_d^{ij}\bar{d}_i q_j H_1 + \bY_e^{ij}\bar{e}_i l_j H_1 
\ee
where $H_{1,2}$ are the Higgs doublets, with the
vacuum expectation values (VEVs)
$v_{1,2}$ breaking the electroweak symmetry,
$(v_1^2 +v_2^2)^{1/2} =v=174$ GeV. 
The $3\times 3$ Yukawa matrices $\bY_{u,d,e}$ are not constrained 
by any symmetry property and thus remain 
arbitrary.\footnote{
The phenomenologically dangerous R-violating terms 
can be suppressed by R-parity. 
In other terms, one can impose the matter parity $Z_2$ 
under which the matter superfields change the sign 
while the Higgs ones are invariant. } 

The second aspect of the flavour problem, specific of SUSY, 
questions the sfermion mass and mixing pattern which is 
determined by the soft SUSY breaking (SSB) terms. 
These include trilinear A-terms:   
\be{A-terms}
\cL_A = \bA_u^{ij}\tilde{\bar{u}}_i \tilde{q}_j H_2 + 
\bA_d^{ij}\tilde{\bar{d}}_i \tilde{q}_j H_1 + 
\bA_e^{ij}\tilde{\bar{e}}_i \tilde{l}_j H_1 
\ee
(the tilde labels sfermions) and soft mass terms:  
\be{soft-m}
\cL_m = \sum_f \tilde{f}_i^\dagger 
(\bmu^2_{\tilde f})^{ij} \tilde{f}_j , ~~~~~~ 
(f= q,\bar{u},\bar{d},l,\bar{e}) 
\ee 
where $\bA_{u,d,e}$ and $\bmu^2_{\tilde f}$ are $3\times 3$ 
matrices with dimensional parameters. 
Theoretical arguments based on the Higgs mass stability 
imply that the typical mass scale $\tilde{m}$ of these terms 
should be of order 100 GeV, maybe up to TeV. The SSB terms 
have no {\it a priori} relation with the Yukawa constants 
$\bY_{u,d,e}$. Hence, one  expects that the splitting 
between the sfermion mass eigenstates be large, of order 
$\tilde{m}$, and in addition the sfermion mixing angles 
controlling the coupling with  fermions and neutral gauginos 
be also large. 
This situation gives rise to dramatic contributions 
to FC  and CP-violating processes.  
For example, the decay rate $\mu\to e+ \gamma$ or 
the CP-violating parameters $\eps_K$ and $\eps^\prime_K$ in 
$K^0-\ov{K}^0$ system would much exceed the experimental 
bounds unless $\tilde{m}$ is larger than 10-100 TeV. 
In this case, however, the advantage of supersymmetry in 
stabilizing the Higgs mass would be lost. Thus, 
experimental limits on  FC processes impose severe 
constraints on the mass and mixing pattern  of the yet 
undiscovered squarks and sleptons. 

As far as neutrinos are concerned, there is no 
renormalizable term that can generate  their masses. 
However, the Majorana masses of neutrinos can emerge 
from the lepton-number violating higher order operator
cutoff by some large scale $M$, e.g. the grand
unification or Planck scale \cite{BEG}: 
\be{Yuk-nu}
\frac{1}{M}\, \bY_\nu^{ij} l_i l_j H_2^2 ~. 
~~~~~  \bY_\nu = \bY_\nu^T
\ee  
Any known mechanism for the neutrino masses reduces to
this effective operator. E.g., in the `seesaw' scheme
\cite{seesaw} it is obtained after integrating out
heavy-neutral fermions with  Majorana masses $\sim M$.
Hence, modulo  Yukawa coupling constants, 
the charged fermion masses are $\sim v$ while 
the neutrino masses are $\sim v^2/M$
which makes it clear why the latter are so small.
However, the matrix $\bY_\nu$ remains arbitrary.

The concept of the grand unification provides more 
constraints on the Yukawa matrices and in this way 
opens up some possibility for predictive schemes of 
quark and lepton masses. 
In the $SU(5)$ model all fermion states 
are unified within 10-plets 
$t_i=(\bar{u},q,\bar{e})_i$ and $\bar5$-plets 
$\bar{f}_i=(\bar{d},l)_i$. 
The minimal structure of the Yukawa terms is the following 
\be{Yuk-su5}
\bG^{ij} \bar{f}_i t_j   \bar{H} +  
\bG^{ij}_u t_i t_j H +  
\frac{1}{M}\bG^{ij}_\nu \bar{f}_i \bar{f}_j H^2   
\ee
where $\bG_u$ and $\bG_\nu$ should be symmetric while 
the form of $\bG$ is not restricted. After the $SU(5)$ 
symmetry breaking, these terms reduce to the standard 
couplings in (\ref{Yuk}) with $\bY_d=\bY_e^T=\bG$. 
This implies that 
the Yukawa eigenvalues are degenerate between the down 
quarks and charged leptons of the same generations. 
Although this prediction for the largest eigenvalues, the  
$b-\tau$ Yukawa unification, is a remarkable success of the 
$SU(5)$ theory, it is completely wrong 
for the light generations.  

The spontaneous breaking of $SU(5)$ to the Standard Model 
by the adjoint superfield $\Phi$ (24-plet)  
can be used to remove that unrealistic degeneracy between 
down-quark and charged leptons. 
The Yukawa coupling matrices can be thought as  operators 
depending on $\Phi$, i.e. 
$\bG=\bG(\Phi)$, $\bG_u=\bG_u(\Phi)$ etc. and hence 
understood as expansion series, e.g. 
\be{Yuk-Phi} 
\bG^{ij}(\Phi) \bar{f}_i t_j  \bar{H}  
= G_0^{ij}  \bar{f}_i t_j   \bar{H} + 
G_1^{ij}\frac{\Phi}{M} t_i \bar{f}_j  \bar{H} + ... 
%\nonumber \\
%+ G_2^{ij}\frac{\Phi^2}{M^2} t_i \bar{f}_j  \bar{H} + ...  
\ee 
where $M$ is some cutoff scale. 
The tensor product $24\times \bar5$ contains both $\bar5$ 
and $\ov{45}$ channels and thus can provide different 
Clebsch factors for the Yukawa entries between the quark 
and lepton states of 
light generations. Clearly, such higher order operators 
can be obtained by integrating out some heavy fermion 
states with masses of order $M$ \cite{FN} just like in 
the (neutrino) seesaw mechanism. 

In this way, the concept of GUT provides a more appealing 
framework for understanding the fermion mass and mixing 
structures. However, at the same time it makes more 
difficult the supersymmetric flavor problem. 
Namely, in the MSSM context natural suppression of the 
flavor-changing phenomena can be achieved by the SSB terms 
universality at the Planck scale, which can be motivated 
in the context of supergravity scenarios \cite{BFS}. 
However, in the SUSY GUT frames this idea becomes 
insufficient -- the physics above the GUT scale does not 
decouple and can strongly violate the SSB terms universality 
at lower scales. In generic SUSY GUTS,  the decoupling of  
heavy states  would  lead to big non-universal 
terms \cite{BH} that  can cause dangerous flavor-changing 
contributions and thus pose a serious challenge to the 
SUSY GUT concept. 

An attractive approach to both flavor problems -- 
fermion and sfermion -- 
is to invoke the idea of horizontal inter-family symmetry. 
Several models based on $U(1)$ or $U(2)$ family symmetries 
have been considered in  literature \cite{u1,u2}. 
However, the chiral $U(3)_H$ or its non-abelian part $SU(3)_H$ 
unifying all fermion generations in horizontal triplets 
\cite{John,su3H,PLB85} 
seems to be the most natural candidate for describing the 
family triplication. In this paper we demonstrate 
its power to provide  a coherent picture for the fermion 
and sfermion masses, to explain the origin of the fermion 
mass spectrum and  mixing structure, and to  naturally 
solve the supersymmetric flavor problem. 

In general, to  construct  realistic scenarios, 
we must require that in the horizontal symmetry limit 
the fermions  remain massless, so that they can 
acquire mass only after the horizontal symmetry breaking. 
In this way, the fermion mass and mixing pattern could reflect 
the VEV pattern of the Higgs scalars leading to the spontaneous 
breaking of the horizontal symmetry. In other words, 
the horizontal symmetry should have chiral character.  
Clearly, the horizontal symmetry $SU(3)_H$ 
can be the appropriate chiral symmetry, 
unlike its $SU(2)_H$ sub-group.  

Observe that, in the limit of massless fermions the 
standard model has a large  global chiral symmetry,  
$U(3)^5 = U(3)_q \times U(3)_{\bar u} \times U(3)_{\bar d} 
\times U(3)_l \times U(3)_{\bar e}$, separately transforming 
the quark and lepton superfields of the three families  
$q_i=(u,d)_i$, $\bar{u}_i$, $\bar{d}_i$, 
$l_i=(\nu,e)_i$ and $\bar{e}_i$ ($i=1,2,3$). 
The Yukawa terms explicitly break this symmetry. 
We can suppose, however, that the Yukawa 
couplings emerge as a result of the spontaneous breaking 
of this symmetry. Such a situation will be considered 
in sect. 2. 
We will show that in this case  we can  
obtain a natural fermion mass hierarchy and mixing 
pattern.\footnote{
In the context of  TeV scale gravity theories with extra 
dimensions \cite{ADD}, 
the chiral flavour symmetries of this type can be 
helpful for suppressing FC phenomena \cite{Gia}.} 

However, the maximal flavor symmetry $U(3)^5$ cannot be implemented  
in the   SUSY GUT context.  
Namely, 
in the $SU(5)$ model the fermions of each generation 
are unified into two multiplets  
$t\sim 10$ and $\bar{f}\sim \bar5$, so that the 
maximal global chiral symmetry reduces to 
$U(3)^2=U(3)_t\times U(3)_{\bar f}$. 
In the $SO(10)$ model all fermions of each family fit 
into one multiplet $\psi \sim 16$, therefore 
the flavor symmetry reduces to $U(3)_H$. 
The latter is still chiral, since now all left handed 
fermions transform as 3 and the right handed ones as 
$\bar3$. Therefore, as far as the GUT framework is 
concerned, it would be most natural to consider the 
horizontal symmetry $U(3)_H$ or even only its 
non-abelian factor $SU(3)_H$. In addition it could 
be the local gauge symmetry emerging from some more 
fundamental theory on the same grounds  
as the GUT symmetry itself. 

We consider $SU(3)_H$ as horizontal symmetry group in 
sect. 3.  
There, the key points  of our discussion can be summarised 
%The winning combination SUSY $+$ GUT$\times SU(3)_H$ 
%acts 
in the following way:  

\noindent
$\bullet$ The spontaneous breaking features of $SU(3)_H$ 
turn  the Yukawa constants of the 
low energy theory (MSSM) into  dynamical degrees of freedom 
and fix the inter-family hierarchy in a  natural way. 
Namely, the third generation becomes heavy 
($Y_t\sim 1$), while the second and first ones become 
lighter by successively increasing powers of small 
parameters. 
We also show how to achieve the desired structure of the
horizontal symmetry breaking VEVs.
% will be discussed in sect. 3. 

\noindent
$\bullet$ The adjoint Higgs (24-plet) which provides the 
gauge symmetry breaking  
$SU(5)\to SU(3)\times SU(2)\times U(1)$,  
should be used in the Yukawa operators 
to remove the unrealistic degeneracy between 
the down-quark and charged leptons.  

\noindent
$\bullet$ Most probably, the horizontal symmetry should 
exhibit the analogous breaking pattern 
$SU(3)_H \to SU(2)_H\times U(1)_H$ 
by the adjoint Higgs (octet), with $SU(2)_H$ acting 
between the light (first and second) fermion generations.  
This could naturally reproduce  the observed pattern of the 
quark and lepton mixings.  
%provides that parameter $b$ as a non-trivial Clebsch coefficient 
%between the 1-2 and 3 generations 
%in the Yukawa matrices of quarks and leptons. 
%

%In the next sections, first  we  provide a general 
%operator analysis 
%for building-up the textures (\ref{Stech}) and second 

\section{ Maximal family symmetry $SU(3)^5$ } 

Let us consider the global flavor symmetry 
$SU(3)^3=SU(3)_q\times SU(3)_{\bar u}\times SU(3)_{\bar d}$: 
of the quark sector. The quark superfields transform as 
\be{quarks}
q_i = \left( \begin{array}{c} u \\ d \end{array} \right)_i  
\sim (3,1,1) ,  ~~~  
\begin{array}{c} \bar{u}_j \sim (1,3,1) , \\ 
   \bar{d}_k \sim (1,1,3) , \end{array}  
\ee 
$i,j,k=1,2,3$ are family indices.  
The quark masses emerge from the effective 
operators \cite{AB}:  
\be{q-op}
\frac{X_u^{ji}}{M} \bar{u}_j q_i H_2 ~ +~ 
\frac{X_d^{ki}}{M} \bar{d}_k q_i H_1  
\ee
where $X_u \sim (\bar3,\bar3,1)$ and $X_d \sim (\bar3,1,\bar3)$   
are the horizontal Higgs superfields in the mixed representations 
of $SU(3)^3$, and $M$ is a cutoff scale (= flavor scale). 
In the context of the renormalizable theory, 
these effective operators can emerge by integrating 
out some extra heavy vector-like matter superfields \cite{FN},  
e.g. the weak isosinglets $U,\bar{U}$ and $D,\bar{D}$ 
having the same color and electric charges as $u,\bar{u}$ and 
$d,\bar{d}$ and transforming in the following 
representations of $SU(3)^3$: 
\be{U-D}
U_i,D_i \sim (3,1,1), ~~~~~
\bar{U}^i, \bar{D}^i \sim (\bar3,1,1) .
\ee
The latter can get masses from the VEV of 
some scalar $\Sigma$, which can be a singlet or 
an octet of $SU(3)_q$, $\Sigma \sim (8,1,1)$. 
On the other hand, they can mix with the light states 
via the superfields $X_{u,d}$ and $H_{1,2}$. 
The relevant superpotential reads: 
\be{q-seesaw}
\bar{u} U X_u + \Sigma U\bar{U} + \bar{U} q H_2 +    
%\nonumber \\
\bar{d} D X_d + \Sigma D\bar{D} + \bar{D} q H_1     
\ee
where order one constants are understood at each coupling. 
After the fields $X_{u,d}$ and $\Sigma$ get large VEVs 
the Yukawa matrices get the form 
\be{m-m}
\mat{0}{\bX_u}{H_2}{\bM_U} , ~~~ \mat{0}{\bX_d}{H_1}{\bM_D}
\ee
where $\bX_{u,d} \sim \langle X_{u,d}\rangle$ and 
$\bM_{U,D} \sim \langle\Sigma\rangle$.  
%$\bM_{U,D}$ 
The the effective operators (\ref{q-op}) emerge 
after integrating out the heavy states in the so called 
seesaw limit $\bX_{u,d} \leq M$. 
Namely, diagonalizing the matrices (\ref{m-m}), 
we see that the states $\bar{u},\bar{U}$ and $\bar{d},\bar{D}$ 
are mixed so that the actual light states which couple to  
$q$ via $H_1$ and $H_2$ become:
\be{real} 
\bar{u}'\simeq \bar{u} + \bX_u \bM_U^{-1}\bar{U}, ~~~
%\nonumber \\ 
\bar{d}'\simeq \bar{d} + \bX_d \bM_D^{-1}\bar{D}  .
\ee
Therefore, the Yukawa constants of the 
Standard Model are nothing but  
\be{Yuk-q} 
\bY_u = \bX_u \bM_U^{-1}, ~~~ \bY_d = \bX_d \bM_D^{-1} .
\ee
The heavy fermion mass matrices  are $SU(3)$ invariant, 
$\bM_{U,D} \sim M$ if $\Sigma$ is a singlet, 
or $\bM_{U,D} \sim M\lambda_8$ 
if $\Sigma$ is an octet with a VEV towards the $\lambda_8$  
generator of $SU(3)_q$. In either case they have rather 
democratic structure and cannot give rise to the fermion 
mass hierarchy.  
Thus, the Yukawa matrices will  reflect 
the form of the horizontal symmetry breaking pattern 
by the VEVs of $X_{u,d}$. 

Let us consider now the VEV pattern 
for the horizontal Higgs superfields. 
In order to be able to write superpotential terms 
for $X_{u,d}$, one has to introduce also the 
superfields in the conjugated representations 
$\bar{X}_u\sim (3,3,1)$ and $\bar{X}_d\sim (3,1,3)$. 
The latter do not couple to the fermion sector 
and just play the role of spectators in the fermion 
mass generation. 
The most general renormalizable superpotential 
does not contain, because of  $SU(3)^3$ symmetry, 
any mixed term between $X_u$ and $X_d$, and so it 
has a separable form $W=W(X_u) + W(X_d)$, where:  
\beqn{sup-ud}
&& W(X_u) = \mu_u X_u\bar{X}_u + X_u^3 + \bar{X}_u^3 ,  
\nonumber \\ 
&& W(X_d) = \mu_d X_d\bar{X}_d + X_d^3 + \bar{X}_d^3 .  
\eeqn
Details about  superpotentials 
of these type are given in Appendix. 
By means of  bi-unitary transformations the VEV 
of $X_u$ can be always chosen in the diagonal form, 
$\bX_u = \bX_u^D = {\rm diag}(x^u_1,x^u_2,x^u_3)$, 
while $\bX_d = \bX_d^D V^\dagger$, $\bX_d=(x^d_1,x^d_2,x^d_3)$, 
where the unitary matrix 
$V$ defines the relative orientation of the two matrices 
$\bX_u$ and $\bX_d$ in the $SU(3)_q$ space and it 
is nothing but the CKM mixing matrix of the quarks, $V=V_{\rm CKM}$. 

It is shown in Appendix that in the supersymmetric limit only  
the product of all three eigenvalues of $X_{u}$ ($X_d)$ 
$x^u_1x^u_2x^u_3$ ($x^d_1x^d_2x^d_3$) can be fixed, but not 
each  single eigenvalue.   
The vacuum degeneracy should be removed by the SSB terms. 
Then for a certain range of the coupling constants
the largest eigenvalues of these fields ($x_3^u, x_3^d$) 
can be of order of the cutoff scale $M$ while the 
others are smaller. 

The above can be interpreted in the following manner. 
With respect to operators like (\ref{q-op}), 
the MSSM Yukawa constants as well as the CKM mixing angles 
become dynamical degrees of freedom. 
In particular, the first operator in (\ref{q-op}) implies 
\be{S-ap} 
\bY_u = {\rm diag} (Y_u,Y_c,Y_t)  
%\matr{Y_u}{0}{0} {0}{Y_c}{0} {0}{0}{Y_t} 
%\sim \frac{\langle S^{ij} \rangle}{M} \!=\! 
\sim\frac{1}{M}{\rm diag} (x^u_1,x^u_2,x^u_3) . 
%({\cal S}_1,{\cal S}_2,{\cal S}_3)  
%\matr{{\cal S}_1}{0}{0} {0}{{\cal S}_2}{0} {0}{0}{{\cal S}_3} . 
%~~~~~ \cS_3 \sim M, ~~ \cS_2\sim \eps M, ~~ 
%\cS_1\sim \eps^2 M,  
\ee
In the exact supersymmetric limit the values of the  
constants $Y_{u,c,t}$ are not fixed -- they have flat 
directions where only their product is fixed 
$Y_u Y_c Y_t = (\mu_u/M)^3$.  However, the SSB 
terms could naturally fix the Yukawa constants 
so that $Y_t\sim 1$. (For related discussion, 
see also ref. \cite{Fabio}.) 
The same is true for the constants $Y_{d,s,b}$. 
In this way, we can naturally obtain the following 
hierarchy for the up and down quark Yukawa eigenvalues:  
\beqn{u-d}
Y_t : Y_c : Y_u \sim 1: \eps_u : \eps_u^2 , ~~~
\eps_u = \mu_u/M 
\nonumber \\  
Y_b : Y_s : Y_d \sim 1: \eps_d : \eps_d^2 , ~~~
\eps_d = \mu_d/M 
\eeqn 
which for $\eps_u \sim 1/200$ and $\eps_d\sim 1/20$ 
well describes the observed spectrum of quark masses.

Now what about  the CKM mixing angles ? 
In the SUSY limit the latter are flat directions  
as far as the superpotential is `separable' as in (\ref{sup-ud}) 
and so the VEV orientation of $X_u$ and $X_d$ 
remains arbitrary. However, also this degeneracy can be 
removed by the SSB terms. In particular, 
we  consider the following effective D-term like operators \cite{AB}: 
\beqn{tot} 
&& \frac{1}{M^2} \int d^4\theta z\bar{z} 
\left[\lambda {\rm Tr}(X^\dagger_uX_u X^\dagger_dX_d) + \right.
\nonumber \\ 
&& ~~
\left. \lambda_1 {\rm Tr}(X^\dagger_uX_u\Sigma^\dagger\Sigma) +    
\lambda_2 {\rm Tr}(X^\dagger_dX_d \Sigma^\dagger\Sigma)\right]    
\eeqn
where $z=\tilde{m}\theta^2$,  
$\bar{z}=\tilde{m}\bar{\theta}^2$ are supersymmetry breaking 
spurions, $\tilde{m}$ being a typical SUSY breaking mass.   
Clearly, such effective operators always emerge in the loop 
corrections after the supersymmetry breaking. 
If $\Sigma$ is a singlet, then the VEVs of  
$\bX_u$ and $\bX_d$ are aligned in the $SU(3)_q$ space 
and thus no CKM mixing can show up. 
However, if $\Sigma$ is an octet of $SU(3)_q$, 
then, for positive $\lambda$'s, there is a 
parameter range for which the VEVs of $X_u$ and $X_d$ are not 
 anymore aligned and so nonzero CKM mixings are generated \cite{AB}.

We have  to remark that in this model the pattern of 
the CKM mixing angles is not related to the hierarchy of 
quark mass eigenvalues, and in general they should be large. 
The 1-2 mixing angle is indeed of order 1, $s_{12}\simeq 0.22$, 
while the 2-3 mixing is small, $s_{23}\simeq 0.04$, for 
which  some fine tuning of the parameters is 
required.\footnote{
Interestingly, then the third mixing angle is predicted 
as $s_{13}\simeq (m_s/m_b)^2(s_{12}/s_{23})$, 
in a good agreement with the experimental data \cite{AB}. 
} 

%$[z\bar{z} {\rm Tr}(X^\dagger_uX_u X^\dagger_dX_d)]_D$, 
%which can fix orientation of the matrices $\bY_u$ 
%and $\bY_d$ only to be parallel or anti-parallel, 
%i.e. $V_{\rm CKM} = 1$.  

We consider now the squark mass and mixing pattern in 
this model. By $SU(3)_H$ symmetry reasons, the soft 
mass terms of the states $q,\bar{u},\bar{d}$ as well 
as those of the heavy states $U,\bar{U},D,\bar{D}$ are  
degenerate between families, while the trilinear A-terms 
 have a  structure proportional to the Yukawa couplings 
(\ref{q-seesaw}). Therefore, after integrating out the heavy 
states, the pattern of the soft mass 
terms (\ref{soft-m}) should be the following. 
The states $q=(u,d)$ do not mix with the heavy fermions, so 
the soft masses of the left-handed squarks 
maintain the $SU(3)_H$ degeneracy 
(at the decoupling scale $M$). As for the states $\bar{u}$ 
and $\bar{d}$, they mix with the heavy states according to 
(\ref{real}), so that the soft mass terms of the right-handed 
squarks should look like: 
\beqn{sq}
&& \bmu^2_{\bar{u}} = \tilde{m}_{1u}^2 +  
\tilde{m}_{2u}^2 \bY_u\bY_u^\dagger +
\tilde{m}_{3u}^2 (\bY_u\bY_u^\dagger)^2  
\nonumber \\ 
&& \bmu^2_{\bar{d}} = \tilde{m}_{1d}^2 +  
\tilde{m}_{2d}^2 \bY_d\bY_d^\dagger +
\tilde{m}_{3d}^2 (\bY_d\bY_d^\dagger)^2  
\eeqn 
where the overall factors are of  order $\tilde{m}^2$. 
Hence, the latter are not degenerate, but they are 
fully aligned to the Yukawa matrices $\bY_u$ and $\bY_d$ 
respectively.  In  similar way, one can 
easily see that the trilinear terms $\bA_{u,d}$ (\ref{A-terms}) 
are also fully aligned to the matrices $\bY_{u,d}$.  
Thus, in this theory no FC contributions 
emerge at the flavor scale $M$. Clearly, the initial 
conditions for the SSB terms are different from the 
universal ones usually adopted in the MSSM. For computing 
the SSB terms at the electroweak scale, the above expressions 
are to  be evolved down by the renormalization group equations. 
However, all FC effects will remain under 
control and the observable FC rates should be of the same 
order as in the "universal" MSSM.

Similar considerations can be applied to the lepton sector, 
for which the maximal chiral flavor symmetry is 
$SU(3)^2 = SU(3)_l\times SU(3)_e$: 
\be{leptons}
l_i = \left( \begin{array}{c} \nu \\ e \end{array} \right)_i  
\sim (3,1) ,  ~~~  
\bar{e}_a \sim (1,3) .
\ee 
Therefore, the effective operators for the 
charged lepton and neutrino masses are 
\be{l-op}
\frac{X_e^{ai}}{M} \bar{e}_a l_i H_1 ~ +~ 
\frac{X_\nu^{ij}}{M^2} l_i l_j H_2^2 , 
\ee
where $X_e \sim (\bar3,\bar3)$ and 
$X_\nu \sim (\bar6,1)$ are horizontal Higgs 
superfields.
% and $M$ is the flavor scale. 
Once more, these operators emerge from the decoupling 
of the heavy charged leptons $E,\bar{E}$ and 
neutral leptons $N,\bar{N}$ (right-handed neutrinos) 
in the following representations of $SU(3)^2$: 
\be{E-N}
E_i,N_i \sim (3,1), ~~~~~
\bar{E}^i, \bar{N}^i \sim (\bar3,1) .
\ee
The relevant superpotential terms read: 
\be{l-seesaw}
\bar{e} E X_e + \bar{E} l H_1 +  \Sigma E\bar{E} +  
%\nonumber \\
  N^2 X_\nu + \bar{N} l H_2 +  \Sigma N\bar{N} 
\ee
where $\Sigma$ is some scalar which can be a singlet or 
octet of $SU(3)_l$. 
%$\Sigma \sim (8,1,1)$. 
Therefore, for the lepton Yukawa matrices we obtain: 
\be{Yuk-l} 
\bY_e = \bX_e \bM_E^{-1} , ~~~ 
\bY_\nu/M = \bM_N^{-1}\bX_\nu \bM_N^{-1}
\ee
where $\bX_{e,\nu} = \langle X_{e,\nu}\rangle$, and 
$\bM_{E,N} \sim \langle\Sigma\rangle$ are the 
mass matrices of the heavy states.   
Once again, the VEV 
of $X_\nu$ can be always chosen in the diagonal form, 
$\bX_\nu = \bX_\nu^D$ 
while $\bX_e = \bX_e^D V_l$, where the unitary matrix 
$V_l$ defines the relative orientation of the two matrices 
$\bX_\nu$ and $\bX_e$ in the $SU(3)_l$ space and it 
is related to the neutrino mixing matrix as 
$V_l=V_{\rm MNS}$. As in the case of quarks, the hierarchy 
between the mass eigenstates can find a natural origin 
in the solution of the Higgs superpotential for $X_\nu$ 
and $X_e$, while the mixing pattern will be fixed by  
SSB terms analogous to (\ref{tot}). 
As we remarked above, the large neutrino mixing angles 
can be obtained in this situation in a rather generic case.

\section{The $SU(5)\times SU(3)_H$ model }
%\subsection{$SU(5)\times SU(3)_H$ operator analysis }

Let us consider now the grand unification case. 
In $SU(5)$ model the fermions of each generation 
are unified into two multiplets, 10-plets 
$t=(\bar{u},q,\bar{e})$ and $\bar5$-plets 
$\bar{f}=(\bar{d},l)$. 
We now consider the horizontal symmetry $SU(3)_H$ 
which unifies the three fermion families as: 
\be{ft}
\bar{f}_i \sim (\bar5,3), ~~~~
t_i \sim (10,3), 
\ee
($i=1,2,3$ is a $SU(3)_H$ index),  
while the Higgs superfields are singlets of $SU(3)_H$, 
$H\sim (5,1)$ and $\bar{H}\sim (\bar5,1)$.

%\footnote{  
%The $SU(3)_H$ symmetry itself does not not prevent the 
%$R$-parity violating couplings 
%$\eps^{ijk}10_i \bar{5}_j \bar{5}_k$ 
%from where the $B-L$ violating terms 
%$u^c d^c d^c,~ q d^c l,~ e^c l l$ arise. 
%For a general discussion on the relation between 
%$R$ parity and horizontal symmetries, see \cite{ZE}. 
%In particular,  in case of the horizontal symmetry 
%$SU(4)_H$ the R-parity could emerge as an 
%accidental symmetry. One has to remark, however, 
%that in many cases considered below the 
%matter parity automatically follows from the 
%discrete symmetries imposed on the model.}  

Since the fermion bilinears transform as 
$3\times 3 = \bar{3} + 6$, their ``standard'' Yukawa 
couplings to the Higgses are forbidden by the horizontal 
symmetry.
%\footnote{
%The following discussion can  equally apply in both the    
%cases of global and local horizontal symmetry, 
%though it would be more appealing  
%to regard $SU(3)_H$ as gauge symmetry, like $SU(5)$.} 
%
Hence, the fermion masses can be induced only by 
higher order operators involving a set of ``horizontal'' 
Higgs superfields $X^{ij}$ in two-index representations 
of $SU(3)_H$: 
symmetric $X_s^{ij}\equiv S^{\{ij\}}\sim (1,\bar6)$ and 
antisymmetric $X_a^{ij}\equiv A^{[ij]} = 
\eps^{ijk} A_k\sim (1,3)$:\footnote{
The theory may also contain  conjugated Higgses 
$\bar{X}_{ij}$ in representations $\bar{S}\sim (1,6)$ 
and $\bar{A}\sim (1,\bar3)$.  
These usually are needed for writing non-trivial 
superpotential terms in order to generate the 
horizontal VEVs (see next Section). 
These fields, however, do not couple to the 
fermions (\ref{ft}) and thus do not 
contribute to their masses.}  
\be{hoo}
\frac{S^{ij}}{M} t_i t_j H + 
\frac{S^{ij}+A^{ij}}{M} \bar{f}_i t_j \bar{H} 
+ \frac{S^{ij}}{M^2} \bar{f}_i \bar{f}_j H^2 
\ee
where $M$ is some large scale (flavor scale). 
In this way, the fermion mass hierarchy can be 
naturally linked to the hierarchy of the horizontal symmetry 
breaking scales \cite{su3H,PLB85}. 
Needless to say, because of the $SU(5)$ symmetry, the  
antisymmetric Higgses $A$ can participate only in second 
term. 

In particular, let us assume that the horizontal Higgses 
include a sextet $S$ and one or more triplets $A$.\footnote{
The alert reader  will notice our improper language:  
we often use for brevity and simplicity  
 `sextet' (or `triplet') also when we deal with the `anti'- 
representation.} 
Without loss of generality, the VEV of $S$ 
can be taken diagonal:
\be{S} 
\langle S^{ij} \rangle = 
\matr{{\cal S}_1}{0}{0} {0}{{\cal S}_2}{0} {0}{0}{{\cal S}_3} , 
~~~ \cS_3 \!\!\gg \!\!\cS_2\!\! \gg \!\!\cS_1 ,  
\ee
while the triplet  $A^{ij} \equiv \eps^{ijk} A_{k}$   
 can in general have the VEVs towards all three components:
\be{A}
\langle A^{ij} \rangle \!=\!  
\matr{0}{\cA_3}{\cA_2} {-\cA_3}{0}{\cA_1} {-\cA_2}{-\cA_1}{0}\! , ~
\cA_1\! \!>\! \cA_2 \!\!> \!\!\cA_3 .   
\ee
%(see Appendix A). 
Therefore, in the low-energy limit the operators (\ref{hoo}) 
reduce to the Yukawa couplings which in terms of 
the dimensionless VEVs 
$\bS=\langle S \rangle/M$ and 
$\bA=\langle A \rangle/M$ read as:    
\be{Stech}
\bY_u = \bS, ~~~ \bY_d,\bY_e^T = \rho\bS + \bA, ~~~  
\bY_\nu = \eta \bS 
\ee
where $\rho$ and $\eta$ are  proportionality coefficients 
related to different coupling constants in (\ref{hoo}). 
This predictive texture, so called Stech ansatz \cite{su3H,Stech},  
is completely excluded on the phenomenological grounds. 
However, its realistic modifications are possible as 
discussed below.   

In view of the renormalizable theory, the operators (\ref{hoo}) 
can be obtained as a result of integrating 
out some heavy fermion states with mass of order $M$. 
It is natural to assume that this scale itself 
emerges from the VEVs of some fields which can be in 
singlet or octet representation of $SU(3)_H$. 
In this case, all terms in the superpotential  are  
trilinear terms, and one can impose a discrete $\cR$ symmetry 
under which all superfields as well as the superpotential 
changes sign. 
In particular, the operators (\ref{hoo}) can be obtained by
integrating out the following heavy states: 
\beqn{FTN} 
&& T^i \sim (10,\bar3), ~~~ \ov{T}_i \sim (\ov{10},3)
\nonumber \\ 
&& \ov{F}^i \sim (\bar5,\bar3), ~~~~ F_i \sim (5,3)
\nonumber \\ 
&& N^i \sim (1,\bar3), ~~~~ \ov{N}_i \sim (1,3) 
\eeqn 
from the following superpotential terms: 
\beqn{sup-FTN} 
&& W_T = t T H + \bar{f} T \bar{H} +  \Sigma T\ov{T} + \ov{T} t S , 
\nonumber \\ 
&& W_F = \bar{f}F A + \Sigma F\ov{F} + \ov{F} t \bar{H}   
\nonumber \\ 
&& W_N= \bar{f} N H + \Sigma N\ov{N} + S \ov{N}^2 .
\eeqn 
If $\Sigma$ is a singlet, then one immediately obtains 
the ansatz (\ref{Stech}). However, we can 
assume that $\Sigma$ contains 
also the $SU(3)_H$ octet with the VEV towards the 
$\lambda_8$ component, and in addition the antisymmetric 
scalars $A$ contain also the adjoint of $SU(5)$, 
$A\sim (24,3)$. 
Moreover, the form of the superpotential (\ref{sup-FTN}) 
can be motivated by some additional symmetries (for example discrete 
symmetries), 
which differently transform $S$ and $A$. 

Let us now consider the superpotential of the horizontal 
Higgses $S$ and $A$. The invariance under 
 additional discrete symmetries, 
can easily force the superpotential 
of these fields to have a `separable'  form, 
$W = W(S)+W(A)$. In addition, the discrete $\cR$ symmetry 
dictates   superpotentials of the form:  
\be{sup-S}
W(S) = Z(\Lambda^2 - S\bar{S}) + Z^3 + S^3 + \bar{S}^3  
\ee
and\footnote{
If there are at least three triplets, terms like $A_1 A_2 A_3$ and 
$\bar{A}_1 \bar{A}_2 \bar{A}_3$ 
are also allowed \cite{AZ-NPB}.}
%while for antisymmetric representations, $X=A$, we have 
\be{sup-A}
W(A) = Z^\prime(\Lambda^2 - A\bar{A}) , 
\ee
where $Z$ and $Z^\prime$ are some singlet superfields. 
Clearly, if $Z$ has a non-zero VEV, 
there is a solution when the 
sextet $S$ has a diagonal VEV with non-zero eigenvalues 
$\langle S\rangle = {\rm diag} (\cS_1,\cS_2,\cS_3)$. 
In this case, the term $ZS\bar{S}$ plays the role of 
the mass term $\mu S\bar{S}$. Similarly,  from the term $W(A)$ 
 the field $A$ gets 
a non-zero VEV which orientation with respect to that of $S$ 
will be determined by the SSB terms pattern. 

We see that $Y_t\sim 1$ implies $\cS_3\sim M$, close 
to cutoff scale,  
which can naturally arise from the Higgs sector. 
Similarly one can  expect that also $Y_{b,\tau}\sim 1$ 
which  would require large $\tan\beta$ regime. 
However, in realistic schemes also moderate  
$\tan\beta$  can be naturally accommodated \cite{AZ-NPB}.  
%In addition, the neutrino operator translates into 
%$Y_3/M_L \sim\cS_3/M^2\sim Y_t/M$. 
 The flavor scale in the theory 
 can be consistently thought to be 
close to the GUT scale $M_G\sim 10^{16}$ GeV.

In conclusion, this theoretical background allows us to motivate 
the following Yukawa matrices \cite{AZ-NPB}:      
\beqn{Stech-1}
& \bY_u = \bS, 
& \bY_d = \rho\bS + \bb^{-1}\bA_d , \nonumber \\ 
& \bY_\nu= \eta \bb^{-1}\bS ,  
& \bY_e^T = \rho\bS  + \bb^{-1}\bA_e ,  
\eeqn 
where 
%$\bS$ is a symmetric matrix which can be taken diagonal, 
%$\rho$ and $\eta$ are some proportionality coefficients, 
$\bA_{e,d}$ are  antisymmetric matrices  with 
24-plet dependent entries inducing 
different Clebsch factors for the down quarks and charged leptons, 
and $\bb  ={\rm Diag}(1,1,b)$, where $b$ is an asymmetry parameter 
induced by the $SU(3)H$ symmetry breaking due to the interplay 
of the singlet and octet VEVs.   
Clearly the above  pattern represents an extension of the 
Stech-like texture considered in \cite{AZ-NPB}. 
%Fritzsch-like `zero-textures' considered in \cite{BR}. 
A careful analysis proves that the above Yukawa pattern  
provides a successful description of fermion masses and mixing angles 
(for details see \cite{AZ-NPB}). 
Alternatively, in another set up, we  could obtain  
a different predictive pattern:   
\beqn{Stech-2}
& \bY_u = \bS, 
& \bY_d = \bb^{-1}(\rho\bS + \bA_d) , \nonumber \\ 
& \bY_\nu= \eta \bb^{-1}\bS ,  
& \bY_e^T = \bb^{-1}(\rho\bS  + \bA_e) ,  
\eeqn 
Both these patterns have a remarkable property. 
Namely, they offer the key  relation  to understand 
the complementary mixing pattern of quarks and leptons.
The origin of this relation in fact can be traced to 
the coincidence of the Yukawa matrices $\bY_d = \bY^T_e$
in the minimal $SU(5)$ theory. In the textures 
(\ref{Stech-1}) and (\ref{Stech-2}) this relation is 
not exact, but is fulfilled with the accuracy of the 
different Clebsch factors in $\bA_d$ and $\bA_e$. 
Explicitly this means that the 2-3 mixing angles in 
the quark and lepton sectors are, respectively:   
\be{mix23} 
\tan\theta_{23}^q \simeq b^{-1/2}\sqrt{{m_s\over m_b}}, ~~~  
\tan\theta_{23}^l \simeq b^{1/2} \sqrt{{m_\mu\over m_\tau}}  
\ee
which can be correctly fixed for  $ b \sim 10$.
%where $b$ is a dimensionless parameter which in the range 
From here  the following product rule 
is obtained \cite{BR}: 
\be{rule23} 
\tan\theta_{23}^q \tan\theta_{23}^l \simeq 
\left(\frac{m_\mu m_s}{m_\tau m_b}\right)^{1/2} .
\ee
This product rule indeed works remarkably well. 
It demonstrates a `see-saw' correspondence between the 
lepton and quark mixing angles and tells us that whenever 
the neutrino mixing is large, $\tan\theta_{23}^l \sim 1$,   
the quark mixing angle comes out small and 
in the correct range, $\tan\theta_{23}^q \sim 0.04$. 

Let us remark, however, that the patterns considered above 
rely on the fact that the sextet $S$ has a VEV with 
non-zero eigenvalues 
$\langle S\rangle = {\rm diag} (\cS_1,\cS_2,\cS_3)$. 
Such a solution of the Higgs superpotential, with the  hierarchy 
$\cS_3 \gg \cS_2 \gg \cS_1$, is indeed possible if the horizontal 
symmetry $SU(3)_H$ is a global symmetry. 
However, this solution disappears  if $SU(3)_H$ is 
a local gauge symmetry, since it is not compatible 
with vanishing gauge $D$-terms of $SU(3)_H$. 

In this case, however, we have to resort to 
the  solution 
$\langle S\rangle = \langle \bar{S}\rangle = 
{\rm diag}(0,0,\cS)$, which is instead compatible with the 
$SU(3)_H$ $D$-term flatness (see Appendix).  
As for triplet fields, the most general VEV pattern is (see Appendix):
\be{A31}
\langle A\rangle = \langle \bar{A}\rangle = 
\matr{0}{\cA_3}{0} {-\cA_3}{0}{\cA_1} {0}{-\cA_1}{0} .
\ee
%{\rm diag}(\cB,0,\cA)$ 
So, we can start from the effective operators 
\beqn{hoo-SA}
\left(\frac{S^{ij}}{M} + \frac{A^{ij}\Phi}{M^2}\right) t_i t_j H ,~
\left(\frac{S^{ij}}{M} + \frac{A^{ij}\Phi}{M^2}\right) 
\bar{f}_i t_j \bar{H} ,&&\nonumber  \\ 
\left
(\frac{S^{ij}}{M^2} +\frac{A^{ij}\Phi}{M^3} 
\right)\bar{f}_i \bar{f}_j H^2 
\phantom{\left(\frac{A^{ij}\Phi}{M^2}\right) \bar{f}_i \bar{f}_j  
}&&
\eeqn 
which  can be induced  by integrating out 
 heavy states from the appropriate Yukawa superpotential.  
In this case  we find  
the following Yukawa texture \cite{ZB}: 
\be{Fritzsch} 
\bY_{f} \!= \!\matr{0}{A_{3f}}{0} {-A_{3f}}{0}{A^\prime_{1f}} 
{0}{-A_{1f}}{S_f} ,  ~~ f=u,d,e,\nu 
\ee 
which resembles the familiar Fritzsch ansatz \cite{Fritzsch}. 
The latter in fact corresponds to the particular case 
$A^\prime_{1f}=A_{1f}$, which can be obtained 
if the there is no $SU(3)_H$ breaking by the octet 
representation, i.e. the scalar $\Sigma$ is a singlet 
\cite{PLB85}.\footnote{Notice  that in 
this case the neutrino mass matrix can have 
only 33 non-zero element $S_\nu$.} 
This situation is now  completely excluded by the 
experimental data. However, for the case 
$A^\prime_{1f} \neq A_{1f}$, one can achieve a very good 
description of the quark and neutrino mass and mixing 
pattern (this is accounted by the 2-3 asymmetry parameter $b$ 
explicitly shown in eq. (\ref{Stech-1}) ).
Notice, that as long as the octet $\Sigma$ partecipates in 
the $SU(3)_H$ breaking with VEV along the $\la_8$ generator 
and  so  $M \propto \unity + \la_8$, 
also the antisymmetric rep. A  contributes to the neutrino mass matrix 
by filling the 23, 32 entries in a symmetric way \cite{BR}.

We can take a more general approach and consider  effective 
operators which also incorporate the $SU(5)$ adjoint Higgs 
$\Phi$:  
\beqn{hoo-phi}
\left(\frac{X_0^{ij}}{M} + \frac{X_1^{ij}\Phi}{M^2}
+ \frac{X_2^{ij}\Phi^2}{M^3}\right) t_i t_j H , \nonumber \\
\left(\frac{X_0^{ij}}{M} + \frac{X_1^{ij}\Phi}{M^2}
+ \frac{X_2^{ij}\Phi^2}{M^3}\right) \bar{f}_i t_j \bar{H} 
%+ \frac{X_s^{ij}}{M^2} \bar{f}_i \bar{f}_j H^2 
\eeqn 
where $X_{0,1,2}$ are the horizontal scalars, which 
can be symmetric or antisymmetric. (Here, for simplicity, we only 
consider the charged fermion sector.)  
This pattern can be motivated by some additional symmetry 
reasons, which differently transform the horizontal Higgses 
$X_{0,1,2}$. 
E.g. one can consider a $Z_3$ symmetry 
acting on the superfields as 
$\Psi \to \Psi \exp(i\frac{2\pi}{3} Q_\Psi)$, and 
take the corresponding charges as $Q(X_0)=0$, $Q(X_1)=1$, 
$Q(X_2)=2$ and $Q(\Phi)=-1$. 

The underlying renormalizable  superpotential can be:    
\beqn{FT}
&& t T H + \bar{f} T \bar{H} + \Sigma T\bar{T} + X_0 \bar{T} t
\nonumber \\ 
&& X_1 \bar{T} T_1 + \Sigma_1 T_1\bar{T}_1 + \Phi \bar{T}_1 t 
\nonumber \\ 
&& X_2 \bar{T}_1 T_2 + \Sigma_2 T_2\bar{T}_2 + \Phi \bar{T}_2 t 
\eeqn
where $\Sigma$, $\Sigma_1$, $\Sigma_2$ are some superfields 
in singlet or octet representations of $SU(3)_H$ 
(in the following, 
singlets will be denoted as $I$ and octet as $\Sigma$), 
with VEVs of order $M$.
These VEVs can emerge from the Higgs superpotential including 
linear terms $\Lambda_k^2 I_k$ for the singlets, with 
$\Lambda_k \sim M$, and all 
possible trilinear terms consistent with the symmetry 
(among which there are also those like $I\Phi^2$ and $\Phi^3$). 
In this way, all singlet and adjoint fields can get 
order $M$ VEVs.\footnote{The linear terms could effectively 
emerge from the trilinear couplings $IQ\bar{Q}$ with the 
fermions $Q$,$\bar{Q}$ belonging to some extra gauge sector in the 
strong coupling regime.} There is also the possibility to 
generate some of these VEVs by means of the anomalous 
$U(1)_A$ symmetry and in this case  the scale $M$ comes out to be  
slightly bigger than the grand unification scale 
$M_G\simeq 10^{16}$ GeV. 

The observed pattern of the fermion masses clearly 
requires that the "leading" horizontal scalar should 
be a $SU(3)_H$ sextet, $X_0=S_0$.  
 As for other fields 
$X_{1,2}$, these can be sextets or triplets. In the latter 
case we  obtain the Fritzsch-like textures 
(\cite{Fritzsch}) already considered above. 
It is interesting to consider also the case when 
these fields are sextets, $S_{1,2}$ and the effective operators read as:
\beqn{hoo-phi-S}
\left(\frac{S_0^{ij}}{M} + \frac{S_1^{ij}\Phi}{M^2}
+ \frac{S_2^{ij}\Phi^2}{M^3}\right) t_i t_j H , \nonumber \\
\left(\frac{S_0^{ij}}{M} + \frac{S_1^{ij}\Phi}{M^2}
+ \frac{S_2^{ij}\Phi^2}{M^3}\right) \bar{f}_i t_j \bar{H} .
%+ \frac{X_s^{ij}}{M^2} \bar{f}_i \bar{f}_j H^2 
\eeqn
     
We now turn to the superpotential of the horizontal 
Higgses. 
Due to the different quantum numbers of $S_{0,1,2}$ with respect to 
additional discrete symmetries 
%one can easily have a situation when 
the superpotential 
of these  fields is `separable' like,  
$W = \sum W(S_k)$, where 
%In particular, for the fields $X_k$ in symmetric 
%representation, $X=S$, the superpotential has a form:  
\be{sup-Sk}
W(S_k) = Z_k(\Lambda^2 - S_k\bar{S}_k) + S_k^3 + \bar{S}_k^3   .
\ee
In this case, each of the VEVs $\langle S_k\rangle$ can 
have only one non-zero eigenvalue. However these VEVs 
can have non-trivial orientations with respect to each other 
in the $SU(3)_H$ space, with generically large angles 
determined by the SSB terms:
\beqn{s-vev}
&&\langle S_0\rangle \propto \bP_0 = {\rm diag}(0,0,1) ,\\ \nonumber
&& \langle S_1\rangle \propto {\bf U}^\dagger_1\bP_0 {\bf U}_1 ,\\ \nonumber
&& \langle S_2\rangle \propto {\bf U}^\dagger_2\bP_0 {\bf U}_2 , 
\eeqn
where  ${\bf U}_{1,2}$ are $SU(3)_H$ unitary matrices reflecting the 
relative orientation. 
Then we obtain  
Yukawa matrices of the following structure: 
\be{P}
\bY_{f} = \bP_0 + \eps_f \bP_1 + \eps_f^2 \bP_2 , ~~~ f=u,d,e 
\ee  
where $P_{0,1,2}$ are rank-1 matrices with $O(1)$ elements  
which can be chosen as\footnote{
Alternatively, the sextets $S_k$ can be regarded 
as `composite' fields obtained 
from tensor products of the triplets 
$A_k$, i.e.   $S^{ij}_k = A^i_k A^j_k$, 
having VEVs $\langle A_0\rangle \propto (0,0,1), ~
\langle A_1\rangle \propto (0,c,s), ~\langle A_2\rangle \propto (x,y,z)$ 
\cite{compo}.}
\beqn{P-form}
&& \bP_0 = (0,0,1)^T\bullet (0,0,1), 
\nonumber \\
&& \bP_1 = (0,c,s)^T\bullet (0,c,s), 
\nonumber \\
&& \bP_2 = (x,y,z)^T\bullet (x,y,z),
\eeqn
and $\eps_f \sim \langle\Phi\rangle/M$ are the Clebsch 
factors projected out from the couplings of $\Phi$ with 
the different types of fermions. Yukawa matrices of such 
structures have been considered in ref. \cite{so10}, 
and earlier, in the context of radiative mass generation 
mechanism, in ref. \cite{Rattazzi}. 

All these considerations not only lead to a general 
understanding of the fermion mass and mixing pattern, 
but  can also lead to the predictive schemes. 
In addition, the schemes arising from  the 
heavy fermion exchanges  (\ref{FT}), 
exhibit the remarkable alignment of the sfermion 
mass matrices to the Yukawa terms, and thus are 
natural as far as the supersymmetric flavor problem 
is concerned \cite{ZB}. 

To conclude, we add the following remark. The horizontal symmetry 
may  guarantee the  R-parity. 
The $SU(3)_H$ symmetry does not work for this, 
though in certain context it could suppress some 
R-violating terms \cite{BBS}. However, the automatic 
R-parity can be achieved in the context of the 
horizontal symmetry $SU(4)_H$ \cite{ZE}.

\section{Appendix: Horizontal VEV structures}

Consider the following superpotential  including  
the superfields $S=S^{ij}$ and $\bar{S}=\bar{S}_{ij}$:   
\be{H-super}
W_S= -\mu S\bar{S} + S^3 + \bar{S}^3 ,  
\ee
where $\mu$ is some mass parameter, 
%$\mu =\langle Y\rangle$ and     
%the explicit combination 
$S^3 \!= \!\frac13 \eps_{ijk}\eps_{abc} S^{ia} S^{jb} S^{kc}$ 
(similarly for $\bar{S}^3$), 
and order one coupling constants are absorbed. 
Observe also that this superpotential is manifestly invariant 
under $Z_3$ symmetry:  
$S \to \exp(i\frac{2\pi}{3}) S$ and  
$\bar{S} \to \exp(-i\frac{2\pi}{3}) \bar{S}$. 
Without loss of generality, the VEV of $S$ can be chosen 
in the diagonal form, 
$\langle S\rangle ={\rm Diag}(\cS_1,\cS_2,\cS_3)$. 
Then the condition of vanishing $F$-terms 
$F_S,F_{\bar{S}}=0$ implies that 
$\langle\bar{S}\rangle$ is also diagonal, 
$\langle\bar{S}\rangle=
{\rm Diag}(\bar{\cS}^1,\bar{\cS}^2,\bar{\cS}^3)$, 
and that  
\be{FS}
\cS_i\cS_j = \mu \eps_{ijk}\bar{\cS}^k, ~~~ 
\bar{\cS}^i\bar{\cS}^j= \mu \eps^{ijk} \cS_k .
\ee 
So, in the exact supersymmetric limit the VEV pattern of $S$ and 
$\bar{S}$ is not fixed unambiguously and there are  
flat directions representing a two-parameter vacuum valley.   
In other words, 
the six equations (\ref{FS}) reduce to four conditions:
\be{FS1}
\cS_1\bar{\cS}^1 = \cS_2\bar{\cS}^2 = \cS_3\bar{\cS}^3 = \mu^2, 
~~~~\cS_1\cS_2\cS_3 = \mu^3 
\ee
while the others are trivially fulfilled 
(e.g. equation $\bar{\cS^1}\bar{\cS^2}\bar{\cS^3} = \mu^3$ 
follows from eqs. (\ref{FS1})). 
%In the above, for simplicity, we put $\sigma=1$.      
Thus, in principle the eigenvalues $\cS_{1,2,3}$ can be 
different from each other, say $\cS_3 > \cS_2 >\cS_1$. 
Then eqs. (\ref{FS}) imply that $\bar{\cS}_{1,2,3}$ should
have an inverse hierarchy, 
$\bar{\cS}^3 < \bar{\cS}^2 < \bar{\cS}^1$. 
More precisely, we have  
$\bar{\cS}^1:\bar{\cS}^2:\bar{\cS}^3=
\cS_1^{-1}:\cS_2^{-1}:\cS_3^{-1}$. 

The flat directions of the VEVs are lifted by the 
soft SUSY breaking D-like terms:\footnote{
The F-like terms $\sim \int d^2\theta zW$ 
are not relevant for the VEV orientation as far as they 
just repeat the holomorphic invariants like 
$S\bar{S}$ and $\det S$ whose values are already fixed 
by the conditions (\ref{FS1}). }  
\beqn{SSB-S}
\cL&\!\!\! =\!\!\!& - \int d^4\theta z\bar{z} 
\left[\alpha {\rm Tr}S^\dagger S + 
\frac{\beta}{M^2} ({\rm Tr}S^\dagger S)^2 +\right. 
\nonumber \\
&& ~~~\left.\frac{\gamma}{M^2} {\rm Tr}(S^\dagger S S^\dagger S) 
+ ...\right] 
%+ [ S \to \bar{S}] 
\eeqn 
having a similar form also for $\bar{S}$.    
Here $z=\tilde{m}\theta^2$, $\bar{z}=\tilde{m}\bar{\theta}^2$ 
are supersymmetry breaking spurions, with $\tilde{m}\sim 1$ TeV.  
The cutoff scale $M$ is taken as the flavor scale, i.e. 
the same as that in the superpotential  (\ref{hoo}),  
and we assume that $M > \mu$. 
%Therefore, for $S_{1,2,3}$ these 
%terms translate explicitly into the following scalar 
%potential: 
%\beqn{SSB-exp}
%\alpha m^2 (|\cS_1|^2 + |\cS_2|^2 + |\cS_3|^2) +   
%\beta \frac{m^2}{M^2} (|\cS_1|^2 + \\ \nonumber
%|\cS_2|^2 + |\cS_3|^2)^2 +   
%\gamma \frac{m^2}{M^2} (|\cS_1|^4 + |\cS_2|^4 + |\cS_3|^4) + 
%[|\cS_k| \to |\bar{\cS}_k|]   
%\eeqn 
The stability of the scalar potential associated with (\ref{SSB-S}) 
 implies that 
$\beta >0$ and $\gamma > -\beta$, whereas $\alpha$ can be 
positive or negative. In the former case the minimization of 
the potential, under the conditions (\ref{FS1}), 
would imply that $\cS_1=\cS_2=\cS_3=\mu$, i.e. no hierarchy 
between the fermion families.  
In the latter case, however, the largest eigenvalue of 
$S$ and $\bar{S}$, respectively $\cS_3$ and $\bar{\cS}_1$,  
grow up above the typical VEV size $\mu$ and reach  
values of the order of the cutoff scale $M$: 
\be{VEV3}
\cS_3,\bar{\cS}_1 \approx
\left(\frac{\alpha}{2(\beta+\gamma)}\right)^{1/2} M \sim M  .
\ee
%that is $\cS_3,\bar{\cS}_1 \sim M$. 
Then it follows from (\ref{FS1}) that 
\be{VEV12}
\cS_2,\bar{\cS}_2 = \mu \sim \eps \cS_3 , ~~~~  
\cS_1,\bar{\cS}_3 = \frac{\mu^2}{\cS_3} \sim \eps^2 \cS_3 ,   
\ee 
where $\eps \sim \mu/M$. 

Let us now consider the variant  (\ref{sup-S}) 
of the superpotential (\ref{H-super}).
% when we assume a discrete symmetry ${\cal R}$ under which 
%all superfields in the theory as well as the superpotential changes the sign. % 
%Then the superpotential respecting ${\cal R}$ symmetry 
%should include some gauge singlet $Z$: 
%\be{sup-S}
%W(S,\bar{S}) = 
%\Lambda^2 Z - Z^3 - Z S\bar{S} + S^3 + \bar{S}^3 
%\ee 
%where $Z$ is a gauge singlet. 
%Once again, we can always chose the basis when the VEVs of $S$ 
%and $\bar{S}$ have the diagonal form, 
%$\langle S\rangle ={\rm Diag}(\cS_1,\cS_2,\cS_3)$  
%and $\langle\bar{S}\rangle=
%{\rm Diag}(\bar{\cS}_1,\bar{\cS}_2,\bar{\cS}_3)$.  
%
In this case we have two solutions: 

(i) $\langle Z\rangle \neq 0$: Clearly, in this case 
the $F$-term conditions are the same as in (\ref{FS1}) 
apart of the fact that the mass scale $\mu$ should be 
substituted by the Z's VEV,  $\mu =\langle Z\rangle$. 
The latter is then fixed as $\langle Z\rangle = \Lambda$ 
by the condition $F_Z=0$.
Thus we still have flat directions which will be 
stabilized by the soft SUSY breaking terms in order to 
obtain a hierarchy of the eigenvalues 
$\cS_3 : \cS_2 : \cS_1 \sim 1 : \eps : \eps^2$, where 
$\eps = \Lambda/M$. This solution exists if $SU(3)_H$ 
is a global symmetry, however it is no more valid 
if $SU(3)_H$ is local. The reason is simple: the inverse 
hierarchy of the VEVs $S_{1,2,3}$ and $\bar{S}_{1,2,3}$ 
is not compatible compatible with the $D$-flatness 
condition of the gauge terms $D_a = \sum_n X^\dagger_n T^a X_n$, 
where $T^a$ are $SU(3)_H$ generators, $a=1,..8$, 
unless  $\cS_3 = \cS_2 = \cS_1$. This solution, however, is in contrast 
with the  fermion mass hierarchy.   
\footnote{In principle, we  could appeal 
to some extra "spectator" superfields in different 
representations of $SU(3)_H$ with  VEVs oriented 
so that to 
cancel the contributions of $\langle S\rangle$ 
in the $SU(3)_H$ gauge D-terms. In this way, not very 
appealing though, the hierarchical 
VEV solution could be consistent  also in the case of local $SU(3)_H$.} 

(ii) $\langle Z\rangle = 0$: In this case the condition 
$F_Z=0$ tells us that $\sum S_i\bar{S}^i = \Lambda^2$, while 
the conditions 
$F_{S,\bar{S}}=0$ yield $S_iS_j=0$ and 
$\bar{S}^i\bar{S}^j =0$.  
Therefore, the VEVs of $S,\bar{S}$ can have only one 
non-zero eigenvalue, which can be  e.g.  
$\cS_3$ and $\bar{\cS}^3$, so that 
$\cS_3\bar{\cS}^3 = \Lambda^2$. 
%
%$\langle S\rangle ={\rm Diag}(0,0,\cS_3)$  
%and $\langle\bar{S}\rangle= {\rm Diag}(0,0,\bar{\cS}^3)$, 
%
Clearly, this solution is compatible with the 
$D$-flatness condition. The requirement $D_a=0$ 
simply fixes  $\cS_3=\bar{\cS}_3 = \Lambda$. 

Thus, in the case of local $SU(3)_H$ we 
obtain the non-degenerate solution 
$\langle S\rangle = \langle\bar{S}\rangle= 
\Lambda \cdot {\rm diag}(0,0,1)$. In this case 
the operators like (\ref{hoo}) involving $S$ 
can  only induce the third-generation masses and 
hence $\Lambda \sim M$ is needed, which is quite a  
natural assumption, to obtain  order 
one Yukawa couplings.\footnote{
For the sake of correctness, we recall that for 
$\langle S \rangle$ close to the cutoff scale $M$ 
the see-saw approximation cannot be applied anymore 
to  decouple  the heavy states and then obtain 
the effective Yukawa term. 
To be specific, by  decoupling  the heavy states
$T, \bar{T}$ in the superpotential (\ref{sup-FTN}),  the effective 
up-quark Yukawa  term is $\frac{S/M}{1 + (S/M)^2)} t t H$, which 
represents an {\it exact} result. This also demonstrates that the
 occurrence of $\langle S\rangle \sim M$  
does not injure the predictive power of the theory.}

Now consider the superpotential (\ref{sup-A}) for the anti-symmetric 
Higgs superfields $A$. 
%Let us assume now that the theory contains three triplets 
%superfields $A_n=A_{ni}$, 
%and their partners $\bar{A}_n=\bar{A}_{n}^i$,  $n=1,2,3$.  
%One can incorporate them by the following terms 
%in the superpotential:   
%\be{A-super}
%W(A) =  Z^\prime (A\bar{A} -\Lambda^2) + Z^{\prime 3} 
%\ee
%where order 1 coupling constants are understood and 
%$A_1A_2 A_3 \equiv \eps^{ijk} A_{1i}A_{2j} A_{3k}$.  
%For simplicity, we shall take all masses $\mu_n$ equal, 
%$\mu_{1,2,3}=\mu' < M$. 
%and put for coupling constants $\rho=\rho'=1$.  
%In addition, by assuming that the triplets 
%have $Z_3$ charges different from that of $S$, 
%we do not include terms like $SA^2\equiv S^{ij} A_iA_j$. 
In the exact supersymmetric limit the ground 
state has a continuous degeneracy (flat direction) 
related to unitary transformations $A \to UA$ 
with $U\subset SU(3)_H$. 
In other terms, the superpotential $W=W_S + W_A$ 
has an accidental global symmetry $SU(3)_S\times SU(3)_A$, 
with two $SU(3)$ factors independently transforming the 
horizontal superfields $S$ and $A$.   
 
Similarly to the case of the Higgs fields $S, \bar{S}$, 
for the solution with $\langle Z'\rangle =0$ 
the conditions $F_A,F_{\bar A}=0$ can only fix 
the values of the holomorphic invariant $A\bar{A}=\Lambda^2$, 
while the $D$-term flatness requires  
$\langle A\rangle = \langle \bar{A}\rangle$.    
By unitary transformation 
$A \to UA$ ($U\subset SU(3)_A$), one can choose a basis 
where the VEV of $A$ points towards the first and third 
components. The relative VEV orientation between $S$ and $A$ 
should be fixed from the soft D-like terms:  
%is determined by the SSB terms structures 
\beqn{SSB-SA} 
&& \frac{1}{M^2} \int d^4\theta z\bar{z} 
\left[\alpha' {\rm Tr}(S^\dagger S A^\dagger A) + \right.
\nonumber \\ 
&& ~~~~~~
\left. \beta' {\rm Tr}(S^\dagger S\Sigma^\dagger\Sigma) +    
\gamma' {\rm Tr}(A^\dagger A \Sigma^\dagger\Sigma)\right]  .  
\eeqn
with $z=\tilde{m}\theta^2$, 
$\bar{z}=\tilde{m}\bar{\theta}^2$.
% being supersymmetry breaking spurions.
%
Namely, for $\alpha',\beta',\gamma'$ all positive, 
one can have the triplet VEV structure in eq, (\ref{A31})
with $\cA_1$ and $\cA_3$ both non-zero.
For more details on the horizontal VEV 
structures, see also \cite{BDJC}.

\end{document}